\documentclass[iop,twocolappendix]{emulateapj}
\usepackage{amsmath,mathtools,mathrsfs}
\usepackage[caption=false]{subfig}
\usepackage[T1]{fontenc}

\usepackage{hyperref}
\begin{document}
\title{A Red-Noise Eigenbasis for the Reconstruction of Blobby Images}

\author{Pravita Hallur\altaffilmark{1}, Lia Medeiros\altaffilmark{2,*}, Tod R. Lauer\altaffilmark{3}}
\altaffiltext{*}{NSF Astronomy and Astrophysics Postdoctoral Fellow}
\altaffiltext{1}{Indian Institute of Science Education and Research Mohali. Sector 81, SAS Nagar, Mohali, PO Manauli, Punjab 140306, India}
\altaffiltext{2}{School of Natural Sciences, Institute for Advanced Study, 1 Einstein Drive, Princeton, NJ 08540}
\altaffiltext{3}{NSF's National Optical Infrared Astronomy Research Laboratory, Tucson, AZ 85726}

\begin{abstract}
We demonstrate the use of an eigenbasis that is derived from principal component analysis (PCA) applied on an ensemble of random-noise images that have a ``red'' power spectrum; i.e., a spectrum that decreases smoothly from large to small spatial scales. The pattern of the resulting eigenbasis allows for the reconstruction of images with a broad range of image morphologies. In particular, we show that this general eigenbasis can be used to efficiently reconstruct images that resemble possible astronomical sources for interferometric observations; even though the images in the original ensemble used to generate the PCA basis are significantly different from the astronomical images. We further show that the efficiency and fidelity of the image reconstructions depends only weakly on the particular parameters of the red-noise power spectrum used to generate the ensemble of images.
\end{abstract}

 
\section{Introduction}\label{sec:intro}

Principal component analysis (PCA) has proven to be a powerful method for representing data objects, such as spectra or images (see e.g., \citealt{1999AJ....117.2052C,TurkandPentland1991}). It is particularly useful for the reconstruction of observations that have incomplete or missing data, or that are strongly affected by noise (see e.g., \citealt{2010AJ....140..390B}). With these considerations in mind, \citet{2018ApJ...864....7M} advocated PCA to synthesize Event Horizon Telescope (EHT) images of accreting black holes.  Reconstructing images from the sparse EHT interferometric observations is challenging, as it requires sensible inferences on the behavior of the visibility domain not well-sampled by the observations. PCA appears to offer a promising approach to inferring the intrinsic form of the radio emission close to the black hole horizon within the restrictions enforced by the EHT observations. 

The premise of PCA is that the object being reconstructed is a member of a class of objects (or is largely similar to them), where the characteristics of the class can be defined by members that have complete or high quality observational or simulated data.  PCA starts by constructing a correlation matrix of the data vectors associated with each of the $N$ members of a ``training set" of objects whose features are considered to fairly encompass the diversity of the class's observational properties.  An orthogonal basis of eigenvectors is derived from the correlation matrix, which provides an explicit parameterization of where the individual members of the training set fall within the $N-$dimensional space that contains them. A key part of PCA is that since the members of the set are presumed to have similarities in their observational forms, the significant dimensionality of the ``eigenbasis" space, $k,$ is likely to be $<<N.$  PCA, in fact, produces eigenvectors sorted by the amount of variance over the training-set observations that each vector contributes to the total variance over the data. With $k<<N,$ the correlations between the various members of the training set allow for plausible inferences of the form of missing data, or the recognition of noise, in the observations of any additional objects believed to be a member of the class of objects that the training set represents.

In the case of EHT observations of nearby accreting black holes such as Saggitarius A$^*$ (Sgr A$^*$) and Messier 87 (M87), \citet{2018ApJ...864....7M} used a library of high-fidelity high spatial-resolution General Relativisitc MagnetoHydroDynamic (GRMHD) plus radiative transfer and ray-tracing simulations of  radio images of accreting black holes to define the PCA eigenbasis. Although EHT observations of these two objects may not match the particular morphology of the simulated images, the real radio maps for these sources are expected to fall within the space spanned by the simulations. While this reconstruction methodology uses a highly specific eigenbasis, its success is still gauged by general criteria, such as the $\chi^2$-value match between the reconstruction and EHT visibility observations. Stated another way, the PCA methodology does not force a solution, but finds an efficient way to identify a good solution with acceptable $\chi^2.$

In addition to the two main targets for the EHT, Sgr~A$^*$ and M87, the EHT also observes several farther Active Galactic Nuclei (AGN) sources such as 3C279 \citep{2020A&A...640A..69K} and Centaurus A \citep{2021NatAs.tmp..139J}. For these more distant sources the EHT will not be able to reach event horizon scale resolution and the emission structure close to the horizon will not be visible. However, the EHT has achieved remarkable images of the jet structure of 3C279 and Centaurus A.  

The eigenbasis based on simulated images described in \citet{2018ApJ...864....7M} is limited to a specific image morphology, and a different basis would be needed to reconstruct the AGN sources described above; we therefore explore the use of PCA for image reconstruction with an eigenbasis that itself is highly general in form and is derived with no obvious link to the particular morphology of the observations that we seek to model. \citet{2018ApJ...864....7M} began exploring the properties of red-noise eigenbases motivated by the fact that both spatial and temporal variability properties of accretion flows such as Sgr A$^*$ are expected to behave as approximate red-noise power spectra. Both the variability of the observationally measured flux of Sgr A$^*$ (\citealt{2008ApJ...688L..17M,2014MNRAS.442.2797D}) and theoretical simulations (e.g., \citealt{2012ApJ...746L..10D} and \citealt{2015ApJ...812..103C}) have been modeled as red-noise power spectra. Furthermore, interstellar refractive scattering is also expected to affect images of Sgr A$^*$ and can also be approximated as a red-noise process (e.g., \citealt{2016ApJ...826..170J}). 

We return to these red-noise eigenbases as a way to reconstruct real EHT images. Our approach is to construct a PCA eigenbasis from 8196 random ``red-noise" images, where red-noise is a pattern of random fluctuations with a power-spectrum that decreases from large to small spatial scales. In this regard the methodology begins to resemble the type of data modelling that can be done with any number of general-purpose orthogonal bases; however, we start with a structural power-spectrum similar to that of the objects being modeled, and preserve the notion of a compact basis in which the eigenimages remain sorted by the variance they encode in the input training set. Even so, we note that the red-noise approach in the end may not be far removed from the observed morphology of EHT images. At the resolution limit of EHT, the images of AGN reconstructed by other techniques have a smooth ``blobby" appearance. The red-noise training images are also blobby.  In effect we are modeling the observations as a linear combination of the limited set of blobby eigenimages.

This paper is organized as follows. In Section \ref{sec:rednoise} we introduce our red-noise model and in Section \ref{sec:PCA} we discuss the results of applying PCA to red noise images. In Section \ref{sec:projections} we project our PCA eigenbasis onto several images with different morphologies. In Section \ref{sec:parameter} we explore the effect of changing parameters in our noise model on the accuracy of the projections and conclude in Section \ref{sec:conclusion}.

\section{Generating Random Red-Noise Images}\label{sec:rednoise}

\begin{figure}[t]
    \centering
    \includegraphics[width=\columnwidth]{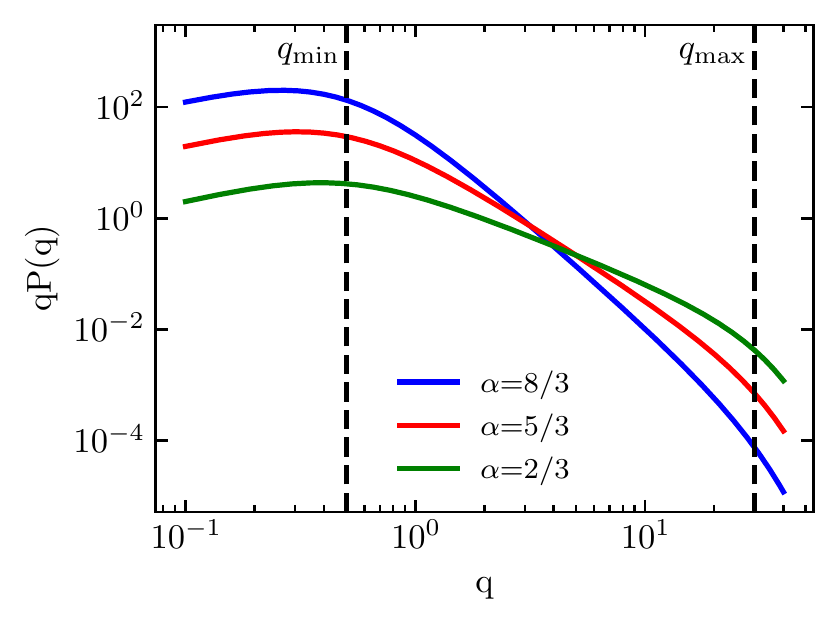}
    \caption{The red-noise power spectrum given by equation \eqref{eq:power_spec} with the parameters $q_{\mathrm{max}}=30$ and $q_{\mathrm{min}}=0.5$ for several values of $\alpha$. Here, $q_{\mathrm{min}}$ and $q_{\mathrm{max}}$ determine the first and second breaks in the spectrum and $\alpha$ determines the slope between $q_{\mathrm{min}}$ and $q_{\mathrm{max}}$.}
    \label{fig:PS}
\end{figure}

\begin{figure*}[htp]
    \centering
    \includegraphics[width=1.1\textwidth]{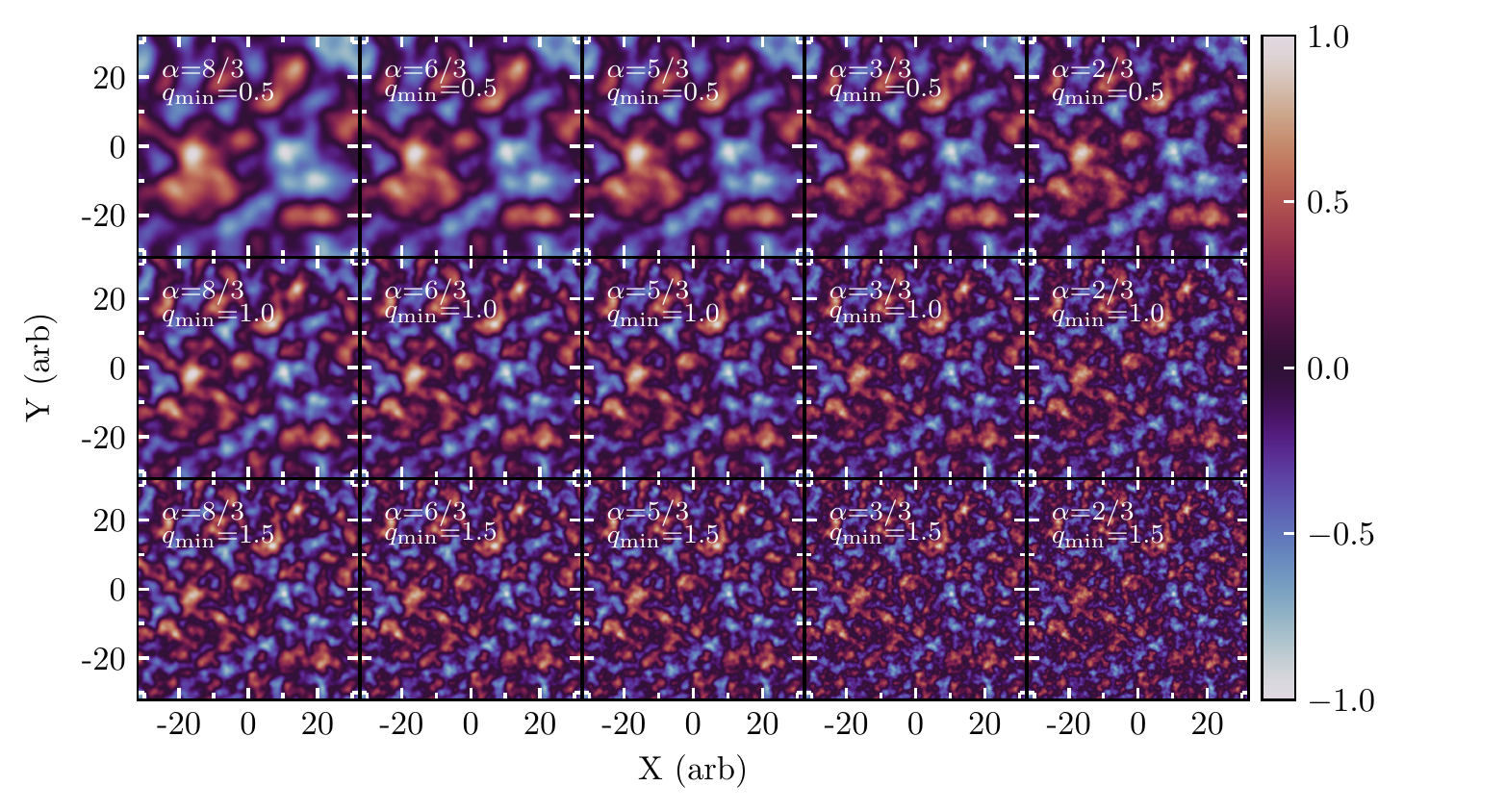}
    \caption{(\textit{top row}) Examples of red-noise images with the power spectrum in Figure \ref{fig:PS} for different values of $\alpha$ (different columns) and $q_{\mathrm{min}}$ (different rows). Increasing the value of $q_{\mathrm{min}}$ or decreasing the value of $\alpha$ both result in finer scale structures. In this figure we use the same random phase fluctuations for all images so that the reader may focus solely on the effect of the changing parameters. However, each image in our ensemble has a different map of random phase fluctuations (see e.g., Figure 14 in \citealt{2018ApJ...864....7M} for an example of red-noise images with different phase fluctuations). }
    \label{fig:RN} 
\end{figure*}

\begin{figure}[h!]
    \centering   
    \includegraphics[width=\columnwidth]{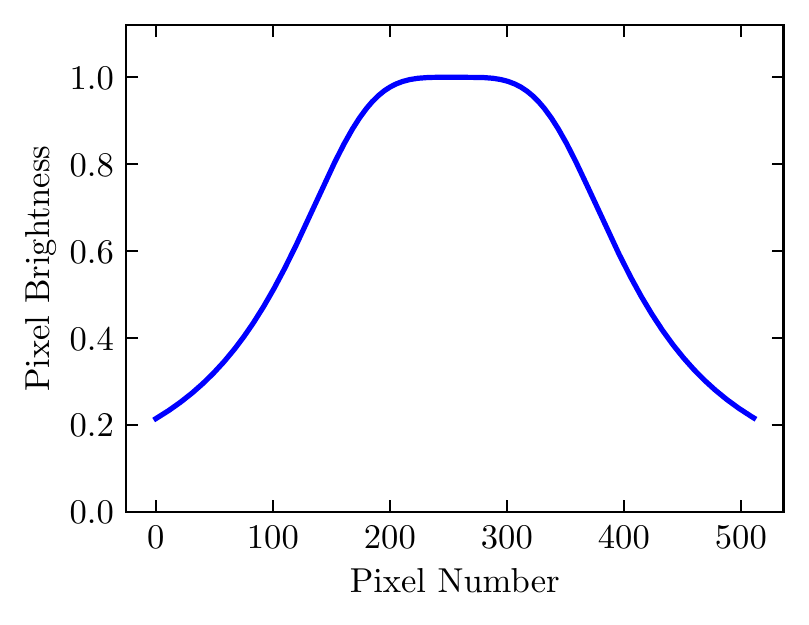}
    \caption{A cross-sectional plot of the Butterworth taper with a radius of $r=120$ pixels and a power law index of $n=2$, which is used to suppress intensity close to the edges of the red-noise images.}
    \label{fig:BWF}
\end{figure}

\begin{figure*}[t]
    \centering
    \includegraphics[width=1.1\textwidth]{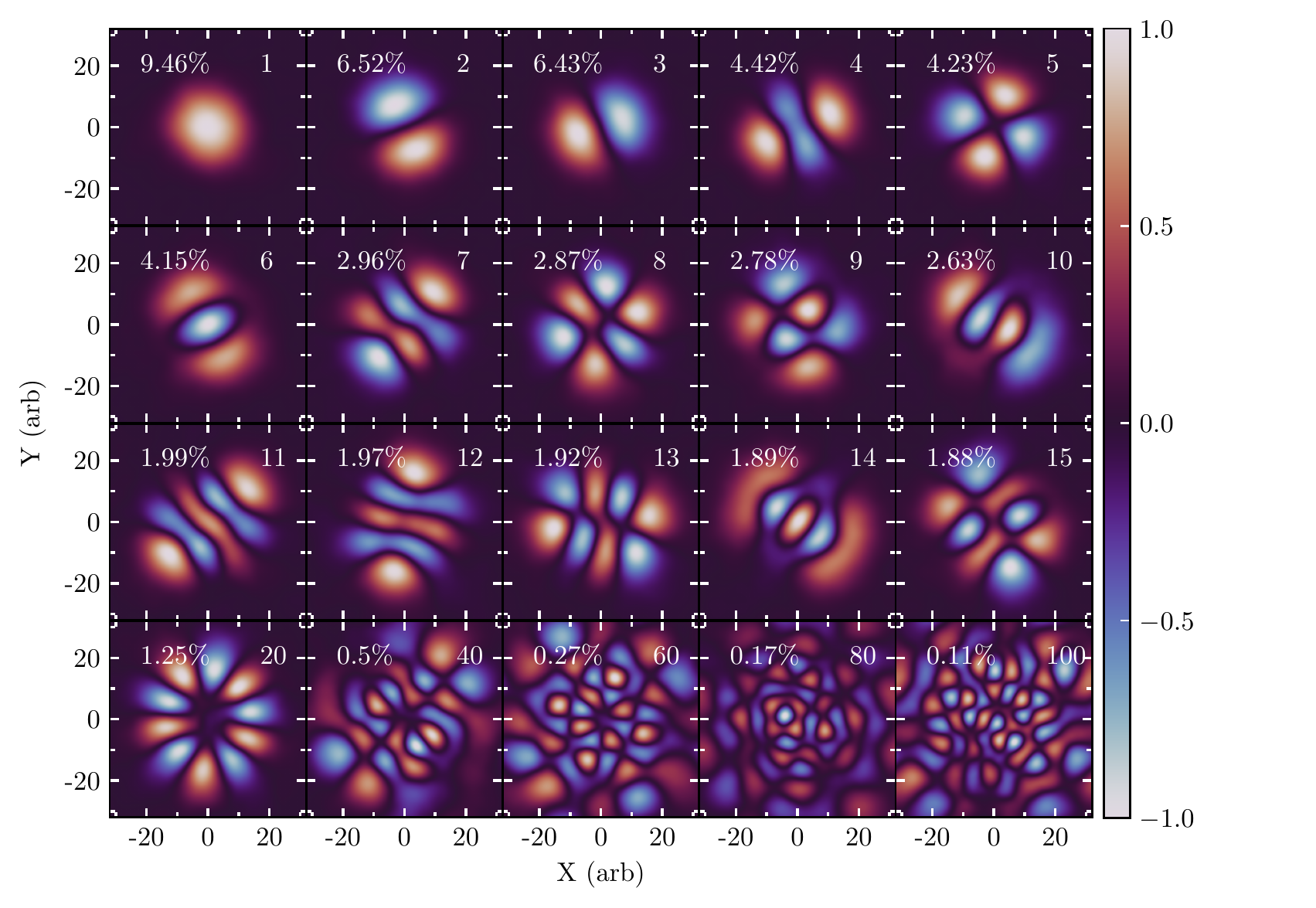}
    \caption{PCA decomposition of a set of red-noise images with $q_{\mathrm{max}}=30$, $q_{\mathrm{min}}=0.5$, and $\alpha=5/3$. The first 15 PCA components are shown in the top three rows and the bottom row shows the 20th, 40th, 60th, 80th, and 100th components. The eigenvalues of the components are displayed at the top left of each panel. The size of the structures in the images get smaller as we increase the component number. The zeroth PCA component is positive definite, and each component has been normalized such that all values are between -1 and 1.}
    \label{fig:PCAcomps}
\end{figure*}

To derive the red-noise eigenbasis we generate an ensemble of random images specified by an input power-spectrum.
Following \citet{2018ApJ...864....7M}, the power spectrum specifies the amplitude at any Fourier frequency, $q,$ as
\begin{equation}\label{eq:power_spec}
    P(q)=2^\alpha\pi\alpha e^{-(q/q_{\mathrm{max}})^2}(q^2+q_{\mathrm{min}}^2)^{-(1+\alpha/2)},
\end{equation}
where $q_{\mathrm{min}}$ and $q_{\mathrm{max}}$ determine the scales of the largest and smallest structures in the images, and $\alpha$ determines the slope of the power spectrum between $q_{\mathrm{min}}$ and $q_{\mathrm{max}}.$  A non-zero $q_{\mathrm{min}}$ also ensures that the flux contained in the image is finite.  The $q_{\mathrm{max}}$ value sets the maximum desired resolution of the images, and prevents the introduction of spurious power on finer scales.  Random images are realized by randomly selected phases for each image, which are drawn from an uniform distribution over 0 to $2\pi.$

In the image domain, the pixel brightness is given by
\begin{equation}\label{eq:2}
    I(\Vec{r})=I_{0}\int d^2qP(q)  \mathrm{exp}[-i\Vec{q}\cdot\Vec{r}]
\end{equation}
where $\Vec{r}$ is the transverse vector on the image plane. We create an ensemble of images with the same power spectrum and random 2D phase distributions.  Figure \ref{fig:PS} shows a few power spectra for various $\alpha$ values. For positive values of $\alpha$, $q_{\mathrm{min}}$ determines the location of the peak of the power spectrum, and therefore sets the size of not only the largest, but also the most common structures in the images.

 Figure \ref{fig:RN} shows example images for different values of $\alpha$ and $q_{\mathrm{min}}$. As expected, the size of the dominant structures in the images scales as $1/q_{\mathrm{min}}$, while changing $\alpha$ determines the relative distribution of different sized structures in the images.

For most astrophysical applications the source emission will be contained close to the center of the image, we therefore smoothly suppress intensity toward the edges of the images by applying a Butterworth taper given by \citep{butterworth1930}
\begin{equation}
    F_\mathrm{{BW}}(b)=\Bigg[1+\bigg(\frac{b}{r}\bigg)^{2n}\Bigg]^{-1/2},
\end{equation}
where $r$ corresponds to the radius and $n$ determines the slope of the cutoff\footnote{We note that throughout the paper the terms ``Butterworth taper'' will be used for a taper in the spatial domain while ``Butterworth filter'' will be used for filtering in the Fourier domain.}. Our images have a resolution of $512\times 512$ and we set $n=2$ and $r=120$ pixels. Figure \ref{fig:BWF} shows a cross sectional plot of the Butterworth taper with these parameters. The structures in our red-noise image ensemble are almost entirely resolved by construction as long as the pixel size is $<< 1/q_{\mathrm{max}}$. 
We do not perform any additional filtering on the noise images since we can constrain the size of the smallest scale structures by setting $q_{\mathrm{max}}$. 

\section{Principal Component Analysis Basis}\label{sec:PCA}

\begin{figure}[h]
    \centering
    \includegraphics[width=\columnwidth]{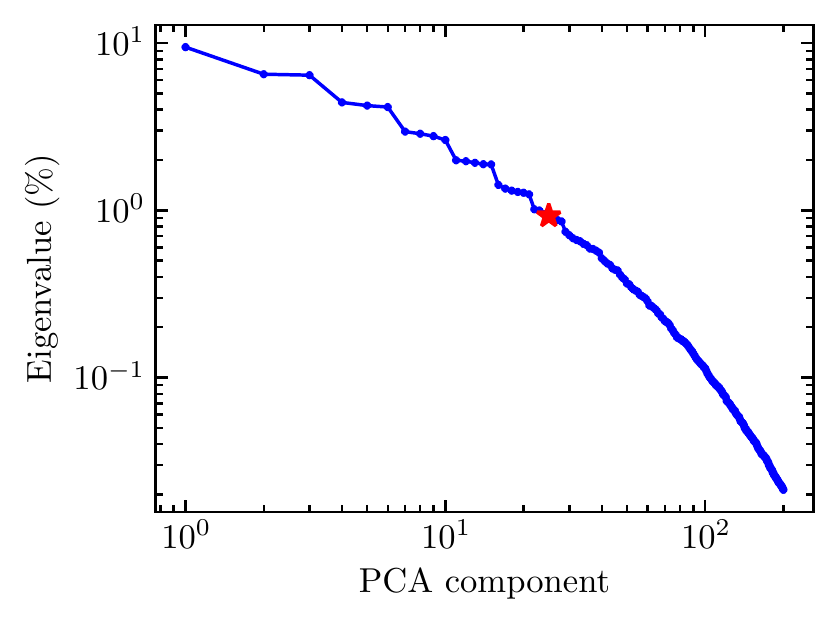}
    \caption{The eigenvalue spectrum of the PCA decomposition shown in Figure~\ref{fig:PCAcomps}. The slope of the eigenvalue spectrum is set by $\alpha$ (see equation~\ref{eq:alpha_gamma}) while the overall normalization is set so that all eigenvalues sum to $100\%$. Although the set of noise images have a low degree of similarity this example spectrum still decreases rather rapidly and only 40 PCA components are needed to explain $80\%$ of the variance in the image ensemble. The red star at 25 PCA components highlights the number of dominant PCA components for a red-noise image ensemble based on $n\approx L^2 q_{\mathrm{min}}^2$.}
    \label{fig:EVS}
\end{figure}

Principal component analysis (PCA) generates an orthogonal basis from the set of random red-noise images by diagonalizing their mutual covariance matrix. Each basis vector, or PCA component, is associated with a particular eigenvalue, which is a measure of the percentage of the variance of the data set that the PCA component explains. The PCA components are ordered by their eigenvalues such that a projection of the original data set onto a lower dimensional space spanned by the first $k$ PCA components will preserve the highest possible percentage of the variance in the data set. This allows us to use PCA to perform efficient dimensionality reduction. In our particular realization that follows we start with a data set of 8192 noise images with the parameters $q_{\mathrm{max}}=30$, $q_{\mathrm{min}}=0.5$, and $\alpha=5/3$. 

Figure~\ref{fig:PCAcomps} shows the first 15 PCA components and the 20th, 40th, 60th, 80th, and 100th component for comparison. The eigenvalues of the PCA components are shown in the top left of each panel. The first few PCA components contain relatively large structures and the structures become progressively smaller for higher components. An analogy can be made with the Fourier transform, the first few components contain information about large scale structures, or low frequencies, while higher order components contain information about smaller scales, or high frequencies. We also note a pattern in the components, the first component divides the region within the taper into one structure, the second and third components divide the region into two structures with a rotation between the second and third component, the fourth and sixth components divide the region into three structures and so on. 

Although all noise images in our data set contain both positive and negative pixels, the first PCA component is positive definite and loosely resembles the taper that was applied to the images, although the radius of the taper is larger than the structure in the first component. When PCA is performed on a set of images with a high degree of similarity, the first PCA component will resemble the average image (see e.g., \citealt{2018ApJ...864....7M}). However, when the set of images does not have a high degree of similarity, the first PCA component will not necessarily resemble the average image. In this case, the structures in our images are random, and therefore have a very low degree of similarity. The only common feature in all images is the Butterworth taper so it is not surprising that the first PCA component resembles this taper.

Since the PCA components are not positive definite, some negative flux in the image reconstructions is possible. We do not constrain our image reconstructions to be positive definite, which results in small amounts of negative flux at the noise level (see \citealt{1981A&A....93..269K} for the effects of constraining images to be positive definite).

Figure~\ref{fig:EVS} shows the eigenvalue spectrum for the PCA decomposition shown in Figure~\ref{fig:PCAcomps}. The ``steps'' present in this eigenvalue spectrum resemble those present in Figure 3 of \citet{2018ApJ...864....7M}, which shows the eigenvalue spectrum of an ensemble of images consisting of a Gaussian moving along a circular path. The patterns present in the PCA components of our red-noise image ensemble discussed above also resemble the patterns of the PCA components of the circular Gaussian model shown in Figure 2 of that paper. As done in the earlier paper, we conclude that the PCA components in each ``step'' in the eigenvalue spectrum are similar, explain a similar amount of variance from the original ensemble, and are frequently just rotated with respect to one another. This step behavior is not noticeable for PCA decompositions of red-noise image ensembles with a smaller number of images and was therefore not noticed in previous work. \citet{2018ApJ...864....7M} showed that the number of dominant noise structures $n$ that can fit in an image of size $L$ is $n\approx L^2q_{\mathrm{min}}^2$, therefore the number of dominant PCA components can also be approximated as $ L^2q_{\mathrm{min}}^2$. For this particular set of noise parameters, the number of dominant PCA components is $\approx 25$, highlighted by a red star in Figure~\ref{fig:EVS}.

\citet{2018ApJ...864....7M} also explored the utility of using PCA to perform dimensionality reduction on GRMHD+ray-tracing simulations of low-luminosity active galactic nuclei (LLAGN), and showed that the slope of the spectrum of eigenvalues of the PCA decomposition of the simulations were diagnostic of the power spectrum of structures in the simulated images. The $\alpha$ parameter in the power spectrum is linearly related to the slope of the eigenvalue spectrum by
\begin{equation}\label{eq:alpha_gamma}
    \gamma\approx \frac{5}{4}\alpha+2
\end{equation}
where $-\gamma$ is the slope of the middle region of the eigenvalue spectrum. As an example, for a spectrum of $\sim30,000$ eigenvalues we use eigenvalues between PCA component 40 and 8,000 to fit for this slope. 

To better understand what range of $\alpha$ would be most appropriate for our ensemble of noise images we use equation \ref{eq:alpha_gamma} and a PCA decomposition of images derived from simulations of accreting black holes. These simulated images are appropriate for the black hole in the center of M87 (with mass $M=6.5\times10^9M_{\odot}$) and span a large range of image morphologies (see \citealt{2021arXiv210503424M} for details on this simulation library and Medeiros et al. 2021b in prep for details on the PCA decomposition of the library). In total, we include 30,720 images from 30 different simulations including both weak and strong magnetic fields with various values for the $R_{\mathrm{high}}$ parameter that parametrizes the electron temperature and various values for the electron number density $n_e$ which affects the thickness of the ring of emission (see also Section~\ref{sec:projections} for a discussion of these parameters and relevant references). For simplicity we kept the black hole spin and the inclination of the black hole's spin axis with respect to the observer's line of sight constant. 

A power-law fit to the middle region of the PCA eigenvalue spectrum of the simulation images returns a value of $\gamma$ of $\sim -3.8$ which results in an $\alpha$ value of $\sim 1.5$. Because of this we explore values of $\alpha$ around 1.5 so that our red noise images have a power spectrum similar to that of simulated images of black hole accretion flows. The value of 1.5 is close to the value of $5/3$ for Kolmogorov turbulence \citep{kolmogorov1941}; however, we emphasize that there is no reason to expect that the power spectrum of image structures should resemble the power spectrum of the turbulence itself, since the structures in the images are affected by not only the turbulence, but also emission mechanisms, and gravitational lensing by the black hole. Still, we explore the $\alpha$ parameter space in terms of fractions of three for consistency with the broader literature on red-noise power spectra that frequently focuses on the Kolmogorov value.

Since the set of noise images have a low degree of similarity, the first eigenvalue is only $\sim 9\%$, which is much smaller than the first eigenvalue of the PCA decomposition of images from a single simulation explored in \citet{2018ApJ...864....7M} which ranged from $\sim66\%$ to $\sim89\%$. The PCA decomposition of the 30,720 images from 30 different simulations described above has a first eigenvalue of $\sim85\%$. Likewise a larger number of PCA components (80 components) are needed to reconstruct $90\%$ of the variance with the PCA derived from the red-noise ensemble than the PCA derived from simulated images (2 PCA components). We note, however, that the magnitude of the first eigenvalue of our red-noise ensemble is not dissimilar to the first eigenvalue of the circular Gaussian model from  \citet{2018ApJ...864....7M}. The magnitude of the first eigenvalue for the PCA decomposition of the red-noise image ensemble did not change significantly when we used 1024 and 4096 images, indicating that the PCA decomposition is only weakly sensitive to the number of images in the ensemble. However, higher numbers of images result in PCA components with a slightly higher degree of symmetry. 

Although the PCA basis derived from simulations is more efficient at reconstructing images that resemble the simulations, our present goal is to explore the utility of using PCA bases derived from noise images as general purpose bases to approximate a broad range of astronomical images. Specifically, we aim to exploit the patterns in the PCA components discussed above to distribute brightness arbitrarily within the taper region. Since the noise images are completely random, the space of images that is spanned by the basis is quite broad. We therefore expect that the basis will be able to approximate a broad range of images, including images that are not within the original image space.

\section{PCA projections}\label{sec:projections}
\begin{figure*}[t]
    \centering
    \includegraphics[width=1.1\textwidth]{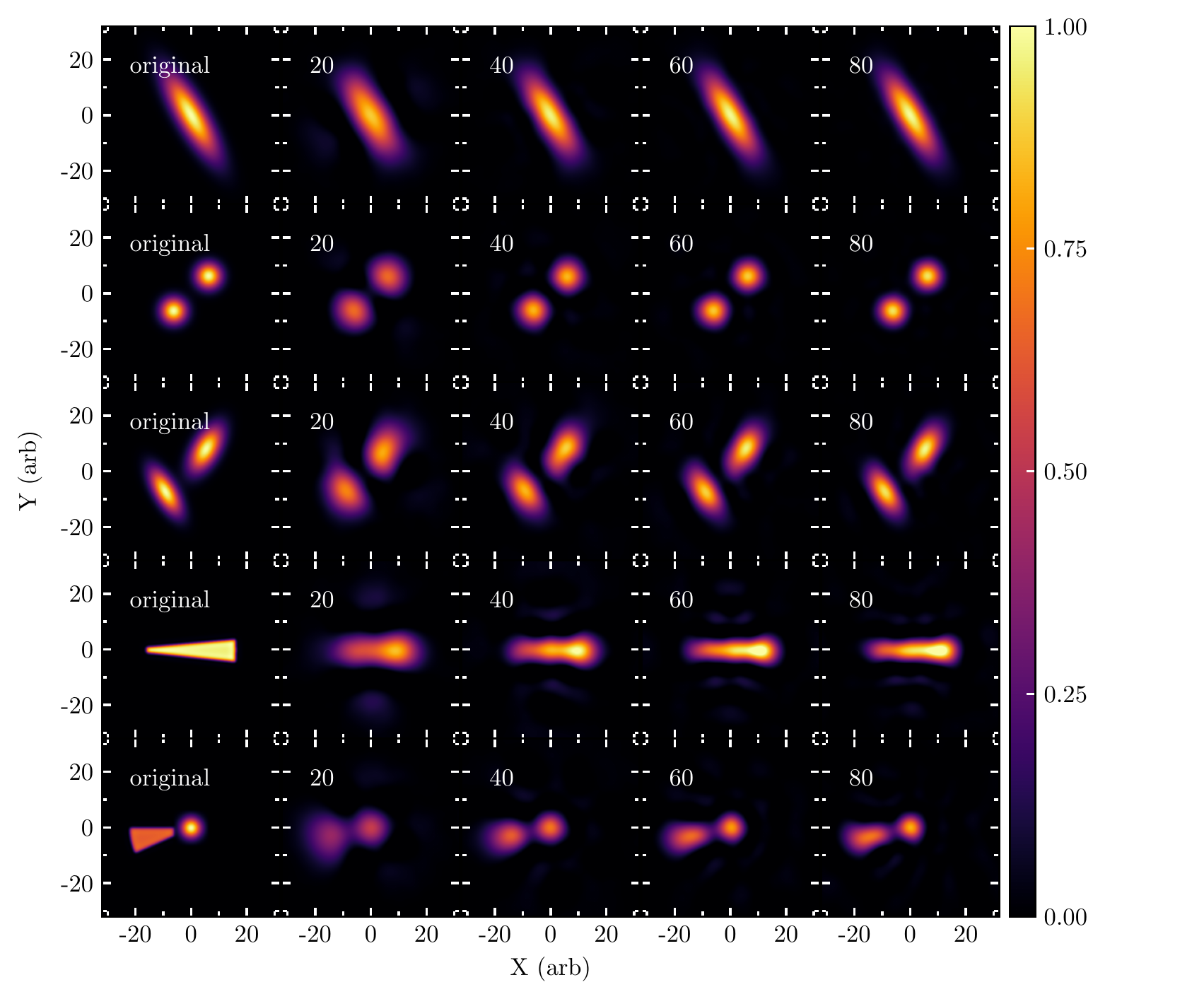}
    \caption{Results of projecting different geometric images onto the PCA basis described in Section~\ref{sec:PCA}. These images are significantly different from the red-noise ensemble of images used to generate the PCA basis. The left most column shows the original images while the second, third, and fourth columns show the results of projecting the original images onto 20, 40, 60, and 80 components \textbf{respectively}. The basic morphology of the simpler images in the top two rows is easily reconstructed by 20 PCA components, however more components are needed to reconstructed the morphology of the bottom three rows. In all cases using more components results in higher image fidelity. Since these are simple geometric shapes we do not ascribe a physical scale to the images and the axes are in arbitrary units. This figure is normalized such that all panels in the same row have the same total flux.}
    \label{fig:PGS}
\end{figure*}

\begin{figure*}[t]
    \centering
    \includegraphics[width=1.1\textwidth]{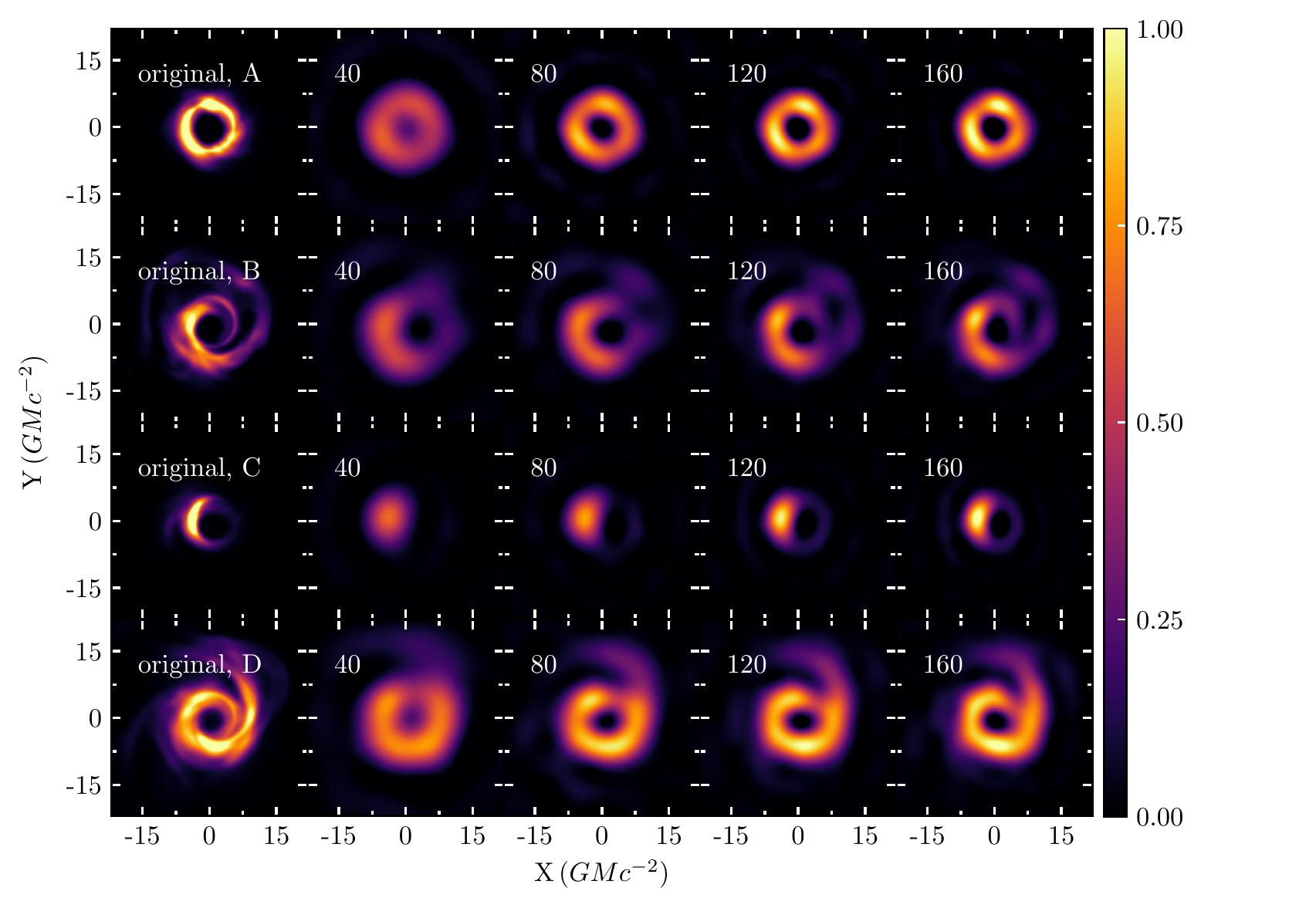}
    \caption{Results of projecting our PCA red-noise basis onto several images resulting from GRMHD + radiative transfer and ray-tracing simulations at 1.3~mm wavelength, the wavelength of EHT observations. The first column shows the original snapshots convolved with a Butterworth filter with a radius of 15$G\lambda$ and a power law index of $n=2$ (see \citealt{2020arXiv200406210P} for details on this filter.) These snapshots are part of the simulation library discussed briefly in Section~\ref{sec:PCA} and originally presented in \citet{2021arXiv210503424M}.  The parameters for each of these snapshots are summarized in Table~\ref{table:GRMHDparams}. The figure is normalized such that all panels in the same row have the same total flux.}
    \label{fig:grid_GRMHD}
\end{figure*}

We project several images onto the PCA basis derived from a red-noise image ensemble discussed in Section~\ref{sec:PCA}. 
We explore how well the first $k$ PCA components of the red-noise basis can reconstruct a broad range of images by projecting the full images onto the PCA components to derive the relative amplitude of each PCA component. The amplitudes are then used to create a linear combination of PCA components that approximates the original image. 

When the method described above is used to reconstruct images that are contained in the original ``training set'' that was used to create the PCA basis, using all of the PCA components in the linear combination will always return the original image. However, in our case we are reconstructing images that are significantly different from the images in the ``training set'', we therefore have no guarantee that we will be able to successfully reconstruct the image even if all of the PCA components are used. Despite this, we still expect to achieve a better approximation to the original image when a larger number of PCA components are used. Ultimately, the number of PCA components that can be fit to a particular interferometric data set will depend on the number of data points in the data set to avoid overfitting. In this section we include the results from projections using a wide range of numbers of PCA components, some of these projections are reasonable for the 2017 EHT data, others are reasonable for future EHT data sets with a more filled array. Although we expect to fit 20-40 PCA components to 2017 EHT data without the need of regularizers, the addition of regularizers would allow for fits with higher numbers of PCA components. We leave an exploration of the effects of sparse interferometric coverage to future work.

In Figure \ref{fig:PGS}, we reconstruct a few images of geometric shapes with varied image morphologies using the PCA basis shown in Figure \ref{fig:PCAcomps}. We show the results from reconstructions using 20, 40, 60, and 80 PCA components. As expected the accuracy of the reconstructions increases with increasing number of components. Specifically, reconstructions with 60 components do a much better job with the small scale details in the images than the reconstructions with 20 components. That being said, the approximate size, shape, and orientation of the elongated Gaussian shown in the top row is well approximated by the reconstruction with only 20 components. Likewise, the locations and approximate shapes of the two Gaussians in the second row are also well approximated by just 20 PCA components, but the size of the Gaussians is better approximated by the reconstruction with 40 components. The two elongated Gaussians in the third row have similar features as the EHT image of 3C279 published in \citet{2020A&A...640A..69K}. The reconstruction with 20 components can accurately reproduce the locations of the two peaks of emission, but cannot reproduce the elongation and orientation of the two Gaussians. The reconstruction with 40 components can reproduce the orientation of the elongation of the two components, but 60 components are needed for an accurate reconstruction of the shape of the two components. 

The wedge feature in the fourth row is meant to approximate a possible elongated jet structure with one end broader than the other. This image and the image in row five have been convolved with a Butterworth filter to remove the sharp edges of the wedges. Although 20 PCA components is enough to reconstruct the relative size and orientation of the wedge, the reconstruction with 40 PCA components achieves a much more accurate representation of the wedge shape. The reconstructions with 60 and 80 components perform even better. Finally, the combination of a Gaussian and a wedge in the last row is significantly harder to reconstruct, but the locations of the two emission peaks are adequately reconstructed with 20 PCA components. Reconstructing the elongation of the wedge structure requires 60 PCA components.

For each of the images in Figure~\ref{fig:PGS} we also calculate the Euclidean distance (see also Mahalanobis distance \citealt{mahalanobis1936}), a measure of the normalized distance between the image and the center of the distribution of the ensemble, which we defined as
\begin{equation}
    d_k = \sqrt{\sum_{n=1}^{N}{\left(\frac{a_{nk}-\bar{a}_{nk}}{\sigma_{a_{nk}}}\right)^2\lambda_{n}}},
\end{equation}
where $a_{nk}$ is the amplitude of the $n-$th PCA component in the projection of the $k-$th image, $\bar{a}_{nk}$ and $\sigma_{a_{nk}}$ are the average and standard deviation of the amplitude of the $n-$th component respectively for all images in the ensemble, and $\lambda_n$ is the eigenvalue of the $n-$th PCA component.The mean of $d_k$ for all red-noise images in the ensemble is $0.99$ while the distances for the five images in Figure~\ref{fig:PGS} range from 75 to 115, indicating that these images are significantly removed from the red-noise images within the ensemble. Even though the example images in Figure~\ref{fig:PGS} are quite different from the overall distribution of red-noise images, the PCA basis is still able to efficiently reconstruct the main image features.

\begin{table}[t]
\begin{center}
 \begin{tabular}{|c c c c c c|} 
 \hline
 Label & B-field Model & Spin & Inclination & $n_{\mathrm{e}}$ & $R_{\mathrm{high}}$ \\ [0.5ex] 
 \hline\hline
 A & MAD & 0.0 & 17 & $5\times10^5$ & 20 \\ 
 \hline
 B & MAD & 0.9 & 17 & $5\times10^5$ & 80 \\ 
 \hline
 C & SANE & 0.9 & 42 & $1\times10^6$ & 80 \\ 
 \hline
 D & MAD & 0.9 & 17 & $1\times10^6$ & 20 \\  [1ex] 
 \hline
\end{tabular}
\caption{Summary of the parameters of the four simulated images shown in Figure~\ref{fig:grid_GRMHD}. See text for definitions of these parameters.}
\label{table:GRMHDparams}
\end{center}
\end{table}
\begin{figure*}[t]
    \centering
    \includegraphics[width=1.1\textwidth]{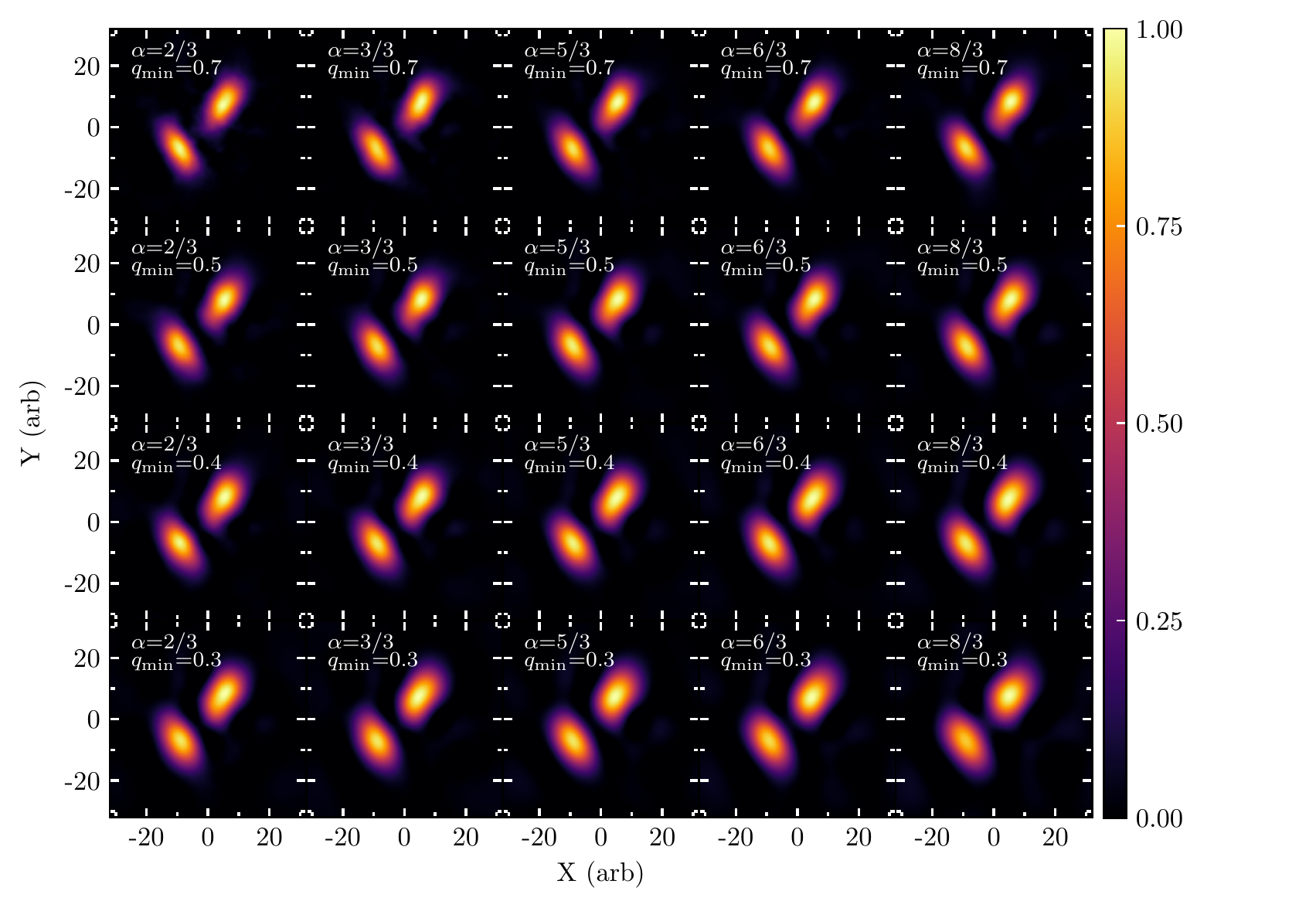}
    \caption{The effects of changing $\alpha$ (different columns) and $q_{\mathrm{min}}$ (different rows) on the projections. All panels show reconstructions with 40 PCA components of the geometric image shown in the third row of Figure \ref{fig:PGS}. The reconstructions depend only weakly on the parameters of the red-noise power spectrum.}
    \label{fig:Param_study_31}
\end{figure*}

Although the focus of this work are distant AGN sources for which the detailed emission structure close to the horizon cannot be resolved, for completeness we also explore the feasibility of reconstructing the near horizon emission structure. The first column of Figure~\ref{fig:grid_GRMHD} shows four representative images from simulations that are appropriate for Saggitarius A$^*$ or Messier 87, which have been filtered with a Butterworth filter to approximate the EHT resolution limit. We include images with a broad range of image morphologies and simulation parameters, both models with strong, ordered magnetic fields (magnetically arrested disk i.e. MAD, see e.g.,  \citealt{2003ApJ...592.1042I}) and models with weak, disordered magnetic fields (standard and normal evolution i.e. SANE, see e.g., \citealt{2012MNRAS.426.3241N}). We also include rapidly spinning and non-spinning black holes, as well as different values for the inclination angle of the observer relative to the black hole's spin axis. The electron number density ($n_e$) and the plasma parameter $R_{\mathrm{high}}$ (see e.g., \citealt{2016A&A...586A..38M, 2019ApJ...875L...4E} for details on $R_{\mathrm{high}}$) are also varied. Table~\ref{table:GRMHDparams} summarizes the parameters for each of the images in Figure~\ref{fig:grid_GRMHD}.

As is evident in Figure~\ref{fig:grid_GRMHD}, the simulated images of near-horizon emission are significantly more challenging to reconstruct with our red-noise basis. In particular, all reconstructions with 40 PCA components are significantly broadened with respect to the filtered snapshot. In contrast, \citet{2018ApJ...864....7M} showed that only 10 PCA components were needed to approximate the simulated images with a PCA basis derived from the simulation itself. However, if we compare the reconstructions with 40 PCA components to the currently published images of M87 (see \citealt{2019ApJ...875L...1E, 2019ApJ...875L...2E,2019ApJ...875L...3E,2019ApJ...875L...4E,2019ApJ...875L...5E,2019ApJ...875L...6E}), we see that the reconstructions with 40 components are already comparable to the resolution of currently published EHT images. In particular, these reconstructions accurately reproduce the approximate size of the emission and can accurately reproduce the brightness depression at the center of the ring images, although there is significant uncertainty in depth of the depression. The reconstructions with 80, 120, and 160 components achieve progressively better reconstructions with increasing fidelity and effective resolution. The Euclidean distances for the images in Figure~\ref{fig:grid_GRMHD} range from 15-30, much smaller than the distances for the images in Figure~\ref{fig:PGS}, but still significantly removed from the red-noise images.

\subsection{Power-Spectrum Parameter Survey}\label{sec:parameter}
Throughout the paper we have used the same PCA basis for all image projections. Here we explore the effects of changing both $\alpha$ and $q_{\mathrm{min}}$ on the fidelity of our image reconstructions. Figure~\ref{fig:Param_study_31} shows reconstructions of a geometric image for various combinations of the $\alpha$ and $q_{\mathrm{min}}$ parameters. Images reconstructed from PCA bases with higher $q_{\mathrm{min}}$ values and/or lower $\alpha$ values result in reconstructions with a ``grainy'' structure while reconstructions with lower $q_{\mathrm{min}}$ and/or higher $\alpha$ result in smoother reconstructions. Our main result, that a red-noise PCA basis can accurately reproduce the main features in the images, is robust to changes in both $q_{\mathrm{min}}$ and $\alpha$, since in all examples we are able to reconstruct the locations and orientations of the two features in the image.

\section{Discussion}\label{sec:conclusion}
We explore the use of PCA eigenbases derived from an ensemble of red-noise images to reconstruct interferometric observations of astronomical sources. This approach is significantly different from what has been commonly done in the PCA literature since dimensionality reduction on noise images is intrinsically difficult and inefficient. However, the particular patterns in the PCA components of red-noise image ensembles divide the possible region of emission into progressively higher numbers of structures, which allows for efficient reconstructions of a broad range of image morphologies. Although some of the PCA components appear quite symmetric, some even resembling spherical Legendre or Zernike polynomials, symmetry properties are not enforced in the PCA components, they are simply optimized to reconstruct the broad range of image morphologies seen in the red-noise image ensemble.

We applied a red-noise generated PCA eigenbasis to the reconstruction of several geometric images that resemble the possible emission structures of AGN jets. We show that our PCA basis can efficiently reconstruct the basic morphology of the images, the approximate size, location, elongation, and orientation of the structures are well approximated with only 20-40 PCA components. Even though these geometric images are quite different from the typical image morphology of the images used to generate the PCA basis, they can be easily reconstructed with a relatively small number of PCA components. With more components, the smaller scale features in the images are also accurately reconstructed. The first EHT observations carried out in 2017 should be able to easily constrain a fit of a linear combination of 20-40 PCA components. However, since 2017 additional telescopes have joined the array possibly allowing for an even greater number of PCA components to be fit to the data. 

We also explore the feasibility of using a red-noise eigenbasis to reconstruct results of GRMHD+radiative transfer and ray-tracing simulations. Although PCA bases derived from the simulations themselves are much more efficient at accurately reconstructing the simulated images with a small number of PCA components, the PCA basis derived from noise is completely agnostic to the underlaying image morphology, and therefore will not bias the reconstructions to approximate GRMHD images. We show that 40-80 PCA components from a red-noise image ensemble is already enough to reconstruct simulated images with a resolution similar to currently published EHT results (see \citealt{2019ApJ...875L...1E, 2019ApJ...875L...2E,2019ApJ...875L...3E,2019ApJ...875L...4E,2019ApJ...875L...5E,2019ApJ...875L...6E}). 

In future work we will include a red-noise PCA basis as part of MARkov Chains for Horizons (\texttt{MARCH}), a Monte Carlo Markov Chain (MCMC) algorithm originally discussed in \citet{2020arXiv200509632P}, and adapted for fitting PCA bases in Medeiros et al. (2021 in prep.). This represents a fundamentally new way of approaching general purpose interferometric image reconstruction of astronomical sources, which can be applied not only to EHT data, but also to observations with other interferometers.

\acknowledgements
P.\;H.\ gratefully acknowledges support from DST-INSPIRE for the INSPIRE scholarship no. MS18198.
L.\;M.\ gratefully acknowledges support from an NSF Astronomy and Astrophysics 
Postdoctoral Fellowship under award no. AST-1903847.


\bibliography{main,my}
\end{document}